\documentstyle[aps,preprint]{revtex}
\begin{document}
\draft
\author{Horacio M. Pastawski\thanks{%
Corresponding author, e-mail: {\it horacio@famaf.fis.uncor.edu}}, Gonzalo
Usaj and Patricia R. Levstein}
\address{Facultad de Matem\'{a}tica, Astronom\'{\i}a y F\'{\i}sica,\\
Universidad Nacional de C\'{o}rdoba, Ciudad Universitaria, \\
5000 C\'{o}rdoba, Argentina}
\title{QUANTUM INTERFERENCE PHENOMENA IN THE LOCAL POLARIZATION DYNAMICS OF
MESOSCOPIC SYSTEMS: AN NMR OBSERVATION.}
\date{\ \ \ \ \ \ \ \ \ \ \ \ \ }
\maketitle

\begin{abstract}
It was predicted that local spin polarization in a ring of five dipolar
coupled spins should present a particular fingerprint of quantum
interferences reflecting both the discrete and finite nature of the system%
{\bf \ [}Phys. Rev. Lett. {\bf 75} (1995) 4310 ]. We report its observation
for the proton system of a (C$_5$H$_5$)$_2$Fe molecule using a rare $^{13}$C
as {\it local probe}{\bf . } Novel high frequency ($\simeq 60k$Hz)
polarization oscillations appear because incomplete $^{13}$C-$^1$H
cross-polarization transfer {\it splits} the polarization state, in a
portion that wanders in the proton system and one that remains in the $^{13}$%
C. They interfere with each other after rejoining.
\end{abstract}


\section{Introduction.}

Mesoscopic systems \cite{Lee} constitute a very active topic because their
transport properties manifest quantum mechanics in its full splendor and
they open new possibilities for the design of devices and finely tailored
systems which started with the tunnel diode\cite{Esaki}. These involve
interference processes both in the space and time domains. The last case
includes the consequences of a finite tunneling time in resonant devices\cite
{Pastawski} and the propagation of a localized excitation in a finite random
media. Here, a simple average over impurities configurations produces an
apparent diffusion, however, a non-linear $\sigma $-model calculation\cite
{Prigodin} predicts strong long time correlations, meaning that a local
excitation returns as a {\it mesoscopic dynamical echo} after diffusing away
and reaching the system boundaries. The observation of these effects in the
time domain is difficult because excitations propagate through the system in
a time scale that is in the frontier of present technological possibilities.
However, a similar phenomenon was also predicted\cite{d-echo} for a local
excitation in a system of nuclear spins with magnetic dipolar interaction,
which evolves with characteristic times on the hundreds of micro-seconds and
can be tested by the powerful NMR spectroscopy. While the discrete nature of
spin systems leads to a fingerprint of quantum interferences in the spin
dynamics of both finite\cite{Braunschweiler} and infinite systems, the
survival of mesoscopic phenomena might not seem obvious since, at high
temperature, different configurations must be averaged and it is often
described as a diffusive process. Therefore, in order to illustrate the
physics we resort to an XY model, where exact analytical results can be
obtained (technical details sketched bellow). Fig. 1 shows the evolution of
the local magnetic polarization $M^I(t)$ for a ring of 21 spins and an
infinite chain at high temperature. At short times a series of quantum
beats, associated with the discrete nature of the system, are developed. At
around 2300$\mu {\rm s}$ it appears a revival of the polarization. A study
of polarization at other sites shows that the wave packet splits in two
parts winding around the ring in opposite directions and the maximum appears
when they meet again in the original site. Similarly, reflections at the
edges of linear chains produce well defined maxima. While we called \cite
{d-echo} this phenomenon {\it quantum dynamical echo} in analogy with the
prediction for quantum dots\cite{Prigodin}, to avoid confusions that might
arise when applied to NMR experiments, we will use {\it mesoscopic beat }%
henceforth. In contrast with ordinary quantum beats, its mesoscopic origin
is shown by a time scale proportional to the system size (see inset of Fig.
1). In dipolar and $J$-coupled systems the range of sizes that present
mesoscopic beats is much more restricted than what the XY example suggests.
For very small systems they become indistinguishable from quantum beats,
while for large systems they are attenuated progressively until they merge
in the background of the other beats. Our finding that local polarization in
a system of five dipolar coupled spins arranged in a ring configuration
should retain quantum beats, was surprising since experiments\cite{ZME}\cite
{Hirschinger} in such a system did not show any signature of these quantum
phenomena. This motivated our present experiments under precisely controlled
conditions.

\section{Theory and experimental methods.}

The NMR radio-frequency pulse sequence developed by Zhang, Meier and Ernst%
\cite{ZME} (ZME), involves cross-polarization (CP) \cite{HH}, to achieve a
two-way transfer of polarization\cite{Levitt} among the spin $S$ of a rare $%
^{13}$C$\,\,\,\,$(1.1\% abundance) and the $I_1$ spin of a directly bonded $%
^1$H\thinspace $\,\,$ . Then, the $^{13}$C can be used as a {\it local probe}
that injects magnetization in the proton and later captures it. The set of
coupled protons within a molecule constitutes the {\it mesoscopic system}
where the spin dynamics can be monitored, while the rest of the crystal
constitutes a 'weakly' interacting reservoir. We measure this spin dynamics
in a polycrystalline sample of ferrocene, (C$_5$H$_5$)$_2$Fe, (Fig.2). Since
the dipolar interaction depends on the angle between the internuclear vector
and the magnetic field, in a general molecule each pair of nuclei can have a
different interaction parameter. Therefore, the resulting Hamiltonian is a
particular case of: 
\begin{equation}
{\cal H}_{II}=\sum_{j>k}\sum_k^{\,\,}d_{j,k}\,\left[ \alpha 2I_j^zI_k^z-%
\frac 12\left( I_j^{+}I_k^{-}+I_j^{-}I_k^{+}\,\right) \right] ,  \label{Hii}
\end{equation}
where subscripts indicate spins. For $\alpha =0,$ it defines an XY model, $%
\alpha =-\frac 12$ describes the $J$-coupling or Heisenberg model, and $%
\alpha =1$ is the truncated dipolar Hamiltonian. In ferrocene at room
temperature, the rings perform very fast rotations around the $C_5$
molecular axis ($\tau _R\approx 10^{-12}{\rm s})$ \cite{Seiler}. In this
last case $d_{j,k}$ are the time-averaged\cite{Slichter} interactions
parameters depending only upon the angle between the rotation axis and the
magnetic field. For each initial state$,\mid i\rangle ,\,$with site $1$
polarized, the probability of finding the same site polarized in the state $%
\langle f\mid $ after a time $t$ is: 
\begin{equation}
P_{f,i}(t)=\left| \langle f\mid \exp [-\frac{{\rm i}}\hbar {\cal H}%
_{II}{}t]\mid i\rangle \right| ^2,  \label{Pt}
\end{equation}
from which the polarization can be calculated summing over all the $N_i$ and 
$N_f$ possible initial and final states:

\begin{equation}
M^I(t)=2\left[ \sum_f^{N_f}\sum_i^{N_i}\frac 1{N_i}P_{f,i}(t)-\frac 12%
\right] .  \label{Mt}
\end{equation}
This is the magnitude which presents quantum mesoscopic beats as function of 
$t$. For the case $\alpha $=0, ( XY-model) an exact mapping\cite{Goncalves}
to a {\it non-interacting} fermions system, allows to sum up the terms in
Eq. (\ref{Mt}) and write the magnetization just as $M^I(t)=\left| \langle
1\mid \exp [-\frac{{\rm i}}\hbar {}{\cal H}_{II}t]\mid 1\rangle \right| ^2,$
the square modulus of a single particle wave function, where state $\mid
1\rangle $ has spin $I_1$ ''up'' and all the others ''down''. Even when the $%
\alpha \neq 0$ cases map to {\it interacting} particles, the essence of this
excitation dynamics is retained until intermediate times.

The best condition to observe quantum beats \cite{d-echo} occurs when the
polarization is quantized along the external magnetic field (laboratory
frame). This is because the neglect of non-secular terms in the dipolar
Hamiltonian is better justified and, since dipolar interactions are
maximized, the time scale of quantum dynamics shrinks minimizing the effects
of other interactions leading to relaxation. The complete pulse sequence is
schematized in Fig. 3. A $\left( \pi /2\right) _x$ pulse on the abundant $^1$%
H spins system creates a polarization that is transferred during $t_C$ to
the rare $^{13}$C system when both are irradiated at their respective
resonant frequencies with field strengths fulfilling the Hartmann-Hahn \cite
{HH} condition $\gamma _IB_{1I}^y=\gamma _SB_{1S}^y\equiv \omega _{1S}^{}$
(here $\omega _{1S}^{}=2\pi \times 44.6{\rm kHz)}.$ After decay of the
proton spin coherence during time $t_S=1{\rm ms}$ the most relevant part
begins: {\bf A) }The magnetization from an initially $y$ polarized $S$ spin
is transferred to the $y$-axis of the $I_1^{}$ spin through a CP pulse of
duration $t_d=85\mu {\rm s}$ (the shortest maximizing the polarization
transfer \cite{Muller} for the selected orientation). {\bf B) }A $\left( \pi
/2\right) _x$ pulse tilts the polarization to the laboratory frame. {\bf C) }%
The $I$-spins evolve freely during a time $t_2$. Thus, the relevant
evolution of the spin-diffusion sequence occurs in the laboratory frame
while $S-$irradiation prevents the system-probe coupling.$\,\,$ {\bf D)} A $%
\left( \pi /2\right) _{-x}$ pulse tilts the polarization back to the
rotating $xy$ plane. {\bf E)} Another CP pulse of length $t_p=t_d$ is
applied to transfer back the polarization to the $x$-axis of $S$. {\bf F) }%
The $S$ polarization is detected while the $I$-system is kept irradiated
(high-resolution condition). An important feature in the pulse sequence of
fig 2 is that while during $t_2$ the spins evolve with Hamiltonian (\ref{Hii}%
), during CP periods $t_d$ and $t_p$ the proton spins evolve with a
Hamiltonian which after truncations is ${\cal H}_{II}^{\prime }=-[\frac 12]%
{\cal H}_{II}^{}$. Thus a part of the free evolution time $t_2$ (around $%
t_m=[\frac 12]$($t_d+t_p)/2=$42.5$\mu {\rm s}$) is spent going backwards in
time (the Loschmidt's daemon\cite{Rhim} is working!) compensating the
undesired evolution. In the ZME experiment an additional polarization
evolution in the rotating frame is allowed by keeping the r.f. in the proton
system during a time $t_1$(i.e. after CP shown in A and before the tilting
pulse B). Therefore a time $t_2=\frac 12t_1+t_m$ is required to reach the
local polarization maximum (polarization echo), after which it evolves
according to Eq. 3.

\section{Results and Discussion.}

In a polycrystal all the angles between the molecular rotation axis and the
magnetic field are equally present. Therefore the $^{13}$C spectrum has a
shape typical of axially symmetric systems with a very well developed peak
corresponding to molecules with their rotational axis lying in the plane
perpendicular to the field. Then, if we select the signal intensity at this
exact frequency, we are monitoring the spin dynamics {\it only} on those
molecules which have the same intramolecular interactions ($d_{1,2}=1576{\rm %
Hz}\times 2\pi \hbar $). Besides, the carrier frequency is set to have this
peak exactly on-resonance avoiding time evolution with chemical shift. The
experimental results with the sequence of Fig 3 are shown in Figure 4 with
circles. Diamonds correspond to a run with the full sequence\cite{ZME} of
ZME with $t_1=80\mu {\rm s.}$ It is presented here as a function of $t_2$
with a shift $-t_1/2$ and properly normalized. The line is obtained by a
spline fitting. For short times we see that superimposed to a parabolic
curve with a maximum (polarization echo) around $\ t_m$ there are high
frequency oscillations. This is a {\it novel quantum phenomenon} that we
discuss bellow. Local polarization decreases until a {\it quantum beat}
appears with a clearly developed maximum at $\approx $370$\mu {\rm s.}$ A
second maximum which, on the basis of numerical analysis, we identify with
the {\it mesoscopic beat,} develops at 520$\mu {\rm s}$ but it is overcome
by an overall attenuation with a characteristic time of about 500$\mu {\rm s}
$. This attenuation contains both the interaction with neighbor molecules
and decoherent processes. The inset shows a sequence of ideal (without
couplings to $^{13}$C) evolutions, $M^I(t)$, calculated with Hamiltonian (%
\ref{Hii}) for a single ring and complete molecule (10 spins in Fig. 2) as
described in \cite{d-echo}. The single ring is the upper dotted curve and is
the case where our previous study of different correlation functions allowed
the identification of the second peak at 580$\mu {\rm s}$ as a mesoscopic
beat \cite{d-echo}. For the molecule we distinguish four situations: rings
rotating independently but keeping a) a staggered configuration , or b) an
eclipsed one ; molecule rotating rigidly with their rings in c) a staggered
configuration or d) an eclipsed one. Each of these gives different sets of
time averaged interaction parameters and represents a progressive increase
in the inter-ring interactions which produces the consequent attenuation of
the interference phenomena. An important conclusion to be drawn from this
sequence is that while changes in the short time evolution, $M^I(t)\simeq 1-%
\frac 12\overline{\Delta \omega ^2}\times t^2,$ are not much apparent, the
effect of those interactions in blurring out the long range interferences is
decisive. This is evident in the decrease of the mesoscopic beat. The above
arguments also lead to interpret the additional attenuation shown up in the
experimental curve as due to intermolecular interactions. Therefore,
although the experimental data can not be fitted to our simulation with
small number of spins, the comparison of these theoretical curves with the
experimental one establishes that the rings rotate independently, a result
compatible with conclusions drawn from proton $T_1$ measurements \cite{Kubo}%
, but can not distinguish between staggered and eclipsed configurations. The
clear correlation between the positions of the experimental and calculated
peaks indicates that this is the first observation of a mesoscopic beat in a
system of spins with magnetic dipolar interactions.

The short time oscillations can be explained as follows: Polarization
initially at the $^{13}$C nuclei has to be transferred to the proton system.
Since systems with more than one proton have an energy uncertainty due to
the inter-proton interaction, it is not possible to obtain an 'exact'
Hartmann-Hahn condition\cite{Levitt}. Therefore, the initial $S$ polarized
state {\it splits} in two: Polarization {\it stays} at $S$ ( state $A)$ with
a probability amplitude $a=\cos \phi $, {\it or} it {\it jumps} to the
proton system (state $B)$ with a probability amplitude $b={\rm i}\sin \phi $%
. During the free evolution time the state $A$ acquires an additional phase
factor due to the spin lock field. When a new CP pulse allows these states
to {\it rejoin}, the phase produces the oscillation in the measured
intensity. The simplest model showing this phenomenon considers a situation
in which the incomplete ($\left| b\right| ^2<1$) CP transfer occurs {\it %
instantaneously}$,$ with both polarization transfers along the $y$%
-direction, and the spin dynamics of only two $^1$H with an XY interaction.
Then the polarization injected in one of them at time $t_2=0$ will be find
again there with a probability amplitude $m_B(t)=\cos [\frac 12\Omega _0t]$,
with $\Omega _0$ obtained as the difference between singlet and triplet
eigenenergies. The ideal magnetization is then $M^I(t)=\left| m_B(t)\right|
^2.$ Because of r.f. irradiation during $t_2$, the $A$ polarization
amplitude is $m_A(t)=\exp \left[ {\rm i}\omega _{1S}\,\,t\right] .$ The
magnetization measured at the $^{13}$C non-ideal probe after a new contact
pulse can be written, after some algebra, as 
\begin{equation}
M^S(t)=\left| b\right| ^4M^I(t)+2\times b^2\cos [\frac 12\Omega _0t]\times
a^2\cos [\omega _{1S}\,t]+\left| a\right| ^4.  \label{Ms}
\end{equation}
Here we identify an interference term proportional to the waiting amplitude, 
$m_B(t)$. To emphasize the effect we re-write this as: $M^S(t)=\left|
b^2m_B(t)+a^2m_A(t)\right| ^2$. The first term represents the probability
amplitude for the $S$ polarization to be transferred to the proton system,
times the probability amplitude for the magnetization to remain in the
original proton site (waiting amplitude), times the probability amplitude to
be transferred back to $S$; the other is the probability amplitude for the
polarization to remain in $S$, evolve and still remain in the $^{13}$C. In
the actual experiment (Fig. 3) the second transfer is to the $x$-axis, this
cancels out the last term in Eq. (\ref{Ms}) and decreases by a factor of 2
the interference term. That gives a coefficient $\left| a\right| ^2/\left|
b\right| ^2$ for the relative importance of the oscillations. From these
considerations, the experimental oscillation amplitude implies that there is
about 90\% efficiency in the polarization transfer from $^{13}$C. This is
also consistent with the experimental lower bound of 85\% efficiency for
each CP transfer obtained from the ratio of intensities measured with $t_2=0$
and with a sequence in which portions A-E have been suppressed. This simple
case has the virtue to present a structure that is maintained for more
complex proton systems: the first and third terms in Eq. (\ref{Ms})
represent the classical effect of sum of probabilities, the second term is
an interference one. In general, the frequencies involved in the first two
terms are not the same. Those contributing to the first have the form $%
\Omega _{ij}=(E_i^N-E_j^N)/\hbar $, where the superscript in the
eigenenergies of Hamiltonian (\ref{Hii}) indicates the number of ''up''
spins in the proton system, while the interference term contains the
frequencies $\Omega _{ij}^{\prime }=(E_i^{N+1}-E_j^N)/\hbar $ i.e. the
subspaces mixed by CP. Hence the expression of the observed polarization $%
M^S(t_2)$ is not simple. A better model takes into account the evolution in
the proton system during the CP periods. Its numerical integration gives, in
agreement with the experiments, both an overall shift in the oscillation
toward frequencies higher than $\omega _{1S}$ and an oscillation maximum
which precedes that of $M^I(t_2-t_m).$ Further comparisons between models
and experiments in simple systems should provide useful information about
decoherence processes and efficiency of cross polarization pulses. On a more
speculative scope, we notice that the described manipulation of polarization
is of the type proposed to implement simple quantum computations \cite
{DiVincenzo} and its further understanding could also contribute to this
developing field.

One might wonder why these wealth of quantum phenomena were not seen in the
ZME experiments performed in a single crystal\cite{ZME}. In order to observe
the interference between polarization pathways it is important to take data
at short enough time intervals. The experiments of ZME were performed at
about the characteristic period (2$\pi /\omega _{1S})$ and that explains why
they could not see this phenomenon. Note that use of higher r.f. power would
make the observation even more elusive. Finally, as mentioned above, in a
polycrystal we are observing {\it simultaneously} all the molecules which
have their rotational axis in the plane perpendicular to the magnetic field;
among them there are some molecules that have the interactions with their
neighbors minimized while other that have them maximized (in a single
crystal this shows up as an orientation dependent second moment $\overline{%
\Delta \omega ^2}$ ). While rings with stronger interaction contribute with
polarization evolutions which do not have quantum beats but decay steadily
much as the plots c$)$ and d$)$ in the inset of Fig. (3), the molecules in
which the characteristic inter-ring vectors form an angle close to the magic
angle can contribute with well defined quantum beats. The importance of the
interaction of neighbor molecules is appreciated in the overall decay of
polarization observed in the experimental data in contrast with the finite
asymptotic polarization in the exact solution of an isolated molecule. While
our experimental data are a superposition of these typical behaviors, the
reported single crystal data, are probably arising from an orientation
favoring intermolecular interactions. Therefore we expect that by choosing
the appropriate orientation in the single crystal a mesoscopic beat better
developed than the one shown here should be observed.

\section{Acknowledgments.}

This work was performed with a Bruker MSL 300 NMR spectrometer at LANAIS de
RMN (UNC-CONICET) with financial support from Fundaci\'{o}n Antorchas,
CONICOR and SeCyT-UNC.

\bigskip\ \ 

{\bf Figure 1: }Evolution of the local magnetic polarization $M^I(t)$ for a
ring of $N=$21 spins at infinite temperature with an XY ($\alpha =0$)
interaction $d=1576$ ${\rm Hz}\times 2\pi \hbar $. Fine line is the square
modulus of the J$_0$ Bessel function, exact solution of an infinite chain.
At short times a series of quantum beats (at around $t_n\simeq z_n\hbar /d$
, $z_n$ the zeroes of J$_1$) are developed which decay with a $t^{-1}$ law,
indicating the excitation propagation. At around 2330$\mu {\rm s}=t_{{\rm %
m.b.}.}$ it appears a revival of the polarization corresponding to the
excitation winding around the ring. The inset shows the mesoscopic beat
time, $t_{{\rm m.b.}},$ for different ring sizes. Notice that $t_{{\rm m.b.}%
}\propto N\times a/v$, where $a$ is a lattice constant and $v\simeq ad/\hbar 
$ an effective group velocity.

\bigskip\ 

{\bf Figure 2:} Ferrocene molecule in a staggered configuration. Fe atom is
at center. Hydrogen atoms are labeled starting from the one which has a rare 
$^{13}$C directly bonded. The eclipsed configuration is obtained rotating
any ring 36$^o$ around the $C_5$ axis.

\bigskip\ 

{\bf Figure 3:} Pulse sequence for proton spin 'diffusion' in the {\it %
laboratory frame}. A $\left( \pi /2\right) _x$ pulse on the abundant $I$
spins creates polarization. This is transferred during $t_C$ to a rare $%
^{13} $C spin $S$. After decay of proton spin coherence during $t_S$:{\bf \
A)}The $S$ magnetization is transferred to the $y$-axis of $I_1^{}$ through
a CP pulse of duration $t_d$. {\bf B)} A $\left( \pi /2\right) _x$ pulse
tilts the polarization to the {\it laboratory frame} where {\bf C) }it
evolves freely during $t_2$. {\bf D)}A $\left( \pi /2\right) _{-x}$ pulse
tilts it back to $y$ and {\bf E)} a CP pulse of length $t_p=t_d$ transfers
the polarization to the $x$-axis of $S$ where {\bf F}) it is recorded.
Dashed and dotted lines show main and secondary probability amplitude
pathways for the polarization.

\bigskip\ 

{\bf Figure 4:} Evolution of a laboratory frame $^1$H spin polarization in
ferrocene as detected in a directly bonded $^{13}$C spin. Notice the maxima
at $\approx $370$\mu $s and $\approx 520\mu $s\ (mesoscopic beat) and
quantum interference oscillations of period 16$\mu $s. {\bf Inset:}
Calculated ideal evolution of local $^1$H polarization in an isolated ring
(dots) and in the complete molecule (see Fig 2) where the rings rotate
independently in: {\bf a)} staggered (full line) or {\bf b)} eclipsed (small
dashed) configurations, and the rigid rotating molecule keeping: {\bf c)}
staggered (long-dashed) or {\bf d)} eclipsed (dot-dashed) rings.

\ 

\end{document}